\documentclass[11pt,a4paper]{article}
\usepackage{amssymb}
\usepackage{times,cite}
\newcommand{\be}{\begin{eqnarray}}
\newcommand{\ee}{\end{eqnarray}}
\newcommand{\bez}{\begin{eqnarray*}}
\newcommand{\eez}{\end{eqnarray*}}
\newcommand{\pa}{\partial}
\newcommand{\la}{\lambda}

\newcommand{\Rf}{\mathfrak{R}}
\newcommand{\A}{\mathcal{A}}
\newcommand{\res}{\mathit{res}}

\newcommand{\id}{\mathrm{id}}

\newcommand{\tpa}{\tilde{\pa}}

\newcommand{\hhi}{\hat{\chi}}
\newcommand{\tE}{\tilde{E}}

\renewcommand{\d}{\mathrm{d}}

\title{\bf Functional representations of integrable hierarchies\thanks{\copyright
       2006 by A. Dimakis and F. M\"uller-Hoissen} }

\author{Aristophanes Dimakis \\
 Department of Financial and Management Engineering, \\
 University of the Aegean, 31 Fostini Str., GR-82100 Chios, Greece \\
 dimakis@aegean.gr
          \and
 Folkert M\"uller-Hoissen \\ Max-Planck-Institute for Dynamics and Self-Organization \\
 Bunsenstrasse 10, D-37073 G\"ottingen, Germany \\
 fmuelle@gwdg.de }

\date{}

\begin{document}

\renewcommand{\theequation} {\arabic{section}.\arabic{equation}}

\newtheorem{theorem}{Theorem}[section]
\newtheorem{lemma}{Lemma}[section]
\newtheorem{proposition}{Proposition}[section]
\newtheorem{definition}{Definition}[section]

\maketitle

\begin{abstract}
We consider a general framework for integrable hierarchies
in Lax form and derive certain universal equations from which
`functional representations' of particular hierarchies
(like KP, discrete KP, mKP, AKNS), i.e. formulations in terms of
functional equations, are systematically and quite easily obtained.
The formalism genuinely applies to hierarchies where the dependent
variables live in a noncommutative (typically matrix) algebra.
The obtained functional representations can be understood as
`noncommutative' analogs of `Fay identities' for the KP hierarchy.
\end{abstract}

\small
\tableofcontents
\normalsize

\section{Introduction}
\setcounter{equation}{0}
In the framework of Gelfand-Dickey-type hierarchies
\cite{Dick03} (see also section~\ref{subsec:GD}),
the commutativity of flows, which is the hierarchy property, is an almost trivial consequence. On the other hand, one is dealing with a rather implicit form of flow equations and it is quite difficult to extract them in more explicit form. In the case where the dependent variables take their values in the (commutative) algebra of functions (of the infinite set of evolution times), expressions of the hierarchy in terms of
(Hirota-Sato) $\tau$-functions can typically be achieved. For example, the
famous KP hierarchy in Gelfand-Dickey-form $L_{t_n} = [(L^n)_{+} , L]$ (see section~\ref{subsec:GD} for notational details) is equivalent to a
`Fay identity' (see \cite{DKJM83,Shio86,Adle+vanM92,Adle+vanM94,Adle+vanM99,Taka+Take95,Dick95,Dick+Stra96,Dick03},
in particular). Such a representation of the hierarchy in the form of
functional equations expresses the complete set of hierarchy equations
directly in terms of the relevant dependent variables as a system of equations
which depend on auxiliary parameters (see also \cite{Nijh+Cape90,Bogd+Kono98,Bogd99,BKM03,Bogd+Kono05,Veks98,Veks02,Prit+Veks02,Veks04} for related work).
\vskip.2cm

In a recent publication \cite{DMH06nahier} we were led to a formula which
may be regarded as a counterpart of the `differential Fay identity' in the case
of the KP hierarchy with variables in a \emph{non}commutative algebra, e.g.,
an algebra of matrices of functions.
In this work we consider the correspondence between such `noncommutative'
(and in particular Gelfand-Dickey-type) hierarchies and equations which
may be regarded as `noncommutative Fay identities' in a quite
general framework.
The main results can actually be proved in a surprisingly general setting.
A more specialized framework then allows to apply the general results
simultaneously in particular to the KP, discrete KP, $q$-KP, AKNS, and
other hierarchies.
\vskip.2cm

In section~\ref{section:hier} we start with a quite general framework for
integrable hierarchies. In subsection~\ref{subsec:GD} we specialize it
to Gelfand-Dickey-type hierarchies and prove a central result of this work.  Section~\ref{section:examples} then concentrates on a more concrete class
of examples. A modified KP hierarchy is treated in section~\ref{section:mKP}.
Finally, section~\ref{section:concl} contains some concluding remarks.

\section{A general framework for hierarchies}
\label{section:hier}
\setcounter{equation}{0}

\subsection{Preliminaries}
Let $\mathbf{t}=(t_1,t_2,t_3,\ldots)$ be a set of independent (commuting)
variables. We introduce
\be
    \chi(\la) := \exp\Big( \sum_{n \geq 1} {\la^n \over n} \, \pa_{t_n} \Big)
               =: \sum_{n \geq 0} \la^n \, \chi_n
    \, , \quad
    \chi(\la)^{-1} =: \sum_{n \geq 0} \la^n \, \hhi_n
\ee
as formal power series in some auxiliary parameter $\la$. Then\footnote{Note
that $\chi_0 = \id = \hhi_0$ and $\chi_1 = \pa_{t_1} = - \hhi_1$.}
\be
    \chi_n = p_n(\tpa) \, , \qquad
    \hhi_n = p_n(-\tpa)  \qquad \quad  n = 0,1,2,\ldots
\ee
where $p_n$, $n=0,1,2,\ldots$, are the elementary Schur polynomials (see
\cite{Macd95,OSTT88}, for example), and
\be
      \tpa := (\pa_{t_1}, \pa_{t_2}/2, \pa_{t_3}/3, \ldots)  \; .  \label{tpa}
\ee
If $F$ depends on $\mathbf{t}$, then
\be
    \chi(\la)(F) = F(\mathbf{t}+[\la]) =: F_{[\la]} \, , \qquad
    \chi(\la)^{-1}(F) = F(\mathbf{t}-[\la]) =: F_{-[\la]}  \, ,
\ee
where
\be
      [\la] := (\la,\la^2/2,\la^3/3,\ldots) \;.
\ee
Using
\be
    {d \over d \la} \chi(\la) = \pa(\la) \chi(\la)   \label{chi_la-ode}
\ee
with the derivation
\be
    \pa(\la) := \sum_{n\geq1} \la^{n-1} \pa_{t_n}  \, ,
\ee
we find
\be
    {d \over d \la} F_{[\la]} = \pa(\la) (F_{[\la]}) \, , \qquad
    {d \over d \la} F_{-[\la]} = - \pa(\la) (F_{-[\la]})  \; .
    \label{dla-id}
\ee
Furthermore, from (\ref{chi_la-ode}) we obtain
\be
    n \, \chi_n = \sum_{k=1}^n \pa_{t_k} \chi_{n-k} \, , \qquad
    n \, \hhi_n = - \sum_{k=1}^n \pa_{t_k} \hhi_{n-k}
    \qquad  n=1,2, \ldots  \; .         \label{chi-Schur-ids}
\ee
Since, as an exponential of a derivation, $\chi(\la)$ is an
automorphism, we have
\be
    \chi(\la)(F G) = \chi(\la)(F) \, \chi(\la)(G)
\ee
on elements $F,G$ of an algebra, the elements of which depend on $\mathbf{t}$.
As a consequence, $\chi_n$ and $\hhi_n$, $n=0,1,2,\ldots$, are Hasse-Schmidt
derivations \cite{Hass+Schm37,Mats86}, i.e. they satisfy the generalized
Leibniz rules
\be
    \chi_n(F G) = \sum_{k=0}^n \chi_k(F) \, \chi_{n-k}(G) \, , \qquad
    \hhi_n(F G) = \sum_{k=0}^n \hhi_k(F) \, \hhi_{n-k}(G) \; .
    \label{chi-HS}
\ee

\subsection{Linear systems and their integrability conditions}
The integrability conditions of the linear system
\be
    \pa_{t_n}(W) = L_n \, W   \qquad \quad n=1,2,\ldots  \label{linsys}
\ee
with invertible\footnote{$W$ should be regarded as a `fundamental matrix
solution' of the linear system.}
$W$ are
\be
    \pa_{t_m}(L_n) - \pa_{t_n}(L_m) = [L_m,L_n]   \label{zero_curv}
\ee
(`zero curvature' or `Zakharov-Shabat' conditions).
Here the $L_n$ and $W$ are elements of a unital algebra $\Rf$.
Let us rewrite the linear system in the following way,\footnote{We may
express the linear system alternatively and equivalently in the form
$\chi_n(W) = H_n \, W$ with $H_n \in \Rf$. Our choice turns out to be
more convenient, however. See the remark in section~\ref{subsec:res}. }
\be
   \hhi_n(W) = E_n \, W \, , \qquad \quad   n=0,1,2,\ldots  \label{linsys-E_n}
\ee
with $E_n \in \Rf$.
Then $E_0 = 1$ (where $1$ stands here for the unit in $\Rf$),
$E_1 = - L_1$,
$E_2 = (1/2)(-L_2 + \pa_{t_1}(L_1) + L_1 L_1)$,
and so forth, so that $E_n$ can be expressed in terms of $L_k$,
$k \leq n$, and their derivatives. Introducing
\be
    E(\la) := \sum_{n \geq 0} \la^n \, E_n \, ,
\ee
the linear system takes the form
\be
    W_{-[\la]} = E(\la) \, W \; .    \label{linsys-HE}
\ee
As a consequence, we have
\be
    W_{-[\la_1]-[\la_2]}
  = E(\la_2)_{-[\la_1]} \, W_{-[\la_1]}
  = E(\la_2)_{-[\la_1]} \, E(\la_1) \, W
\ee
which requires the last expression to be symmetric in $\la_1, \la_2$.
Hence the integrability conditions of the linear system translate to
\be
    E(\la_2)_{-[\la_1]} \, E(\la_1)
  = E(\la_1)_{-[\la_2]} \, E(\la_2)  \; .     \label{EE}
\ee
This formula is of central importance in this work.
Expanding in $\la_1,\la_2$, we obtain the following expression of the
zero curvature conditions,
\be
   \sum_{k=0}^m \hhi_k(E_n) \, E_{m-k} = \sum_{k=0}^n \hhi_k(E_m) \, E_{n-k}
   \qquad \quad m,n=0,1,2, \ldots  \; .
\ee
\vskip.2cm

By use of (\ref{dla-id}) and (\ref{linsys}), we have
\be
    {d \over d \la}(E(\la) \, W)
  = - \pa(\la)(W_{-[\la]})
  = - \pa(\la)(E(\la) \, W)
  = - \pa(\la)(E(\la)) \, W - E(\la) \, L(\la) \, W \, ,
\ee
where
\be
    L(\la) := \sum_{n \geq 1} \la^{n-1} \, L_n \; .
\ee
Hence
\be
    {d \over d \la} E(\la)
  = - \pa(\la)(E(\la)) - E(\la) \, L(\la)  \; .
         \label{dlaE(la)}
\ee
Expansion in powers of $\la$ leads to
\be
    n E_n &=& - \sum_{k=1}^n \Big( \pa_{t_k}(E_{n-k}) + E_{n-k} \, L_k \Big)
    \qquad \quad n=1,2, \ldots \; .    \label{HnEn-recurs}
\ee
This can be used to compute the $E_n$ recursively in terms of the $L_n$
and their derivatives.
\vskip.2cm

\noindent
{\it Remark.} Let $\Theta$ be an automorphism of $\Rf$ which commutes with the
partial derivative operators with respect to the variables $t_n$, and
let us consider an extension of the linear system (\ref{linsys}) of the form
\be
    \Theta(W) = K \, W \; .  \label{Theta-evolution}
\ee
This gives rise to the additional integrability conditions
\be
  \pa_{t_n}(K) = \Theta(L_n) \, K - K \, L_n   \qquad n=1,2,\ldots \, ,
\ee
which can also be expressed as
\be
    K_{-[\la]} \, E(\la) = \Theta(E(\la)) \, K \; .
\ee
If the elements of $\Rf$ depend on a discrete variable,
the shift operator $\Lambda$ with respect to this variable provides us
with an example of such a $\Theta$. Then (\ref{Theta-evolution}) is
a discrete evolution equation.
\hfill $\blacksquare$

\subsection{Lax equations}
Let
\be
     L := W \, \tilde{D} \, W^{-1}     \label{L-dress}
\ee
where $W$ satisfies the linear system (\ref{linsys}) and $\tilde{D} \in \Rf$
is independent of $\mathbf{t}$. This is known as a (Wilson-Sato)
`dressing transformation'.
Differentiation of (\ref{L-dress}) with respect to $t_n$ and use of
(\ref{linsys}) leads to the Lax equations
\be
    \pa_{t_n}(L) = [L_n, L]   \qquad \quad n=1,2, \ldots \; .  \label{Lax-hier}
\ee
Typically we should look for a receipe which determines the $L_n, E_n$
in terms of $L$ (cf section~\ref{subsec:GD}).
An alternative form of equations (\ref{Lax-hier}) is obtained as follows,
\be
    E(\la) \, L \, W
  = E(\la) \, W \, \tilde{D}
  = W_{-[\la]} \, \tilde{D}
  = ( W \, \tilde{D})_{-[\la]}
  = ( L \, W )_{-[\la]}
  = L_{-[\la]} \, W_{-[\la]}
  = L_{-[\la]} \, E(\la) \, W  \, . \quad
\ee
Hence
\be
    L_{-[\la]} \, E(\la) = E(\la) \, L \; .   \label{LE-EL}
\ee
Expanding in $\la$, this becomes
\be
    \hhi_n(L) &=& [E_n,L] - \sum_{k=1}^{n-1} \hhi_k(L) \, E_{n-k}
       \qquad \quad  n=1,2, \ldots \; .
\ee
Clearly, this set of equations is equivalent to (\ref{Lax-hier}).

\subsection{Gelfand-Dickey-type hierarchies}
\label{subsec:GD}
Let $\Rf$ now be a unital associative algebra with a projection $( \, )_{-}$
such that $\Rf = (\Rf)_{-} \oplus (\Rf)_{+}$ where
$( \, )_{+} = \id - ( \, )_{-}$, and $(\Rf)_{-}$, $(\Rf)_{+}$
are subalgebras. Furthermore, we assume that $\Rf$ is generated by
$L$ via the product in $\Rf$ and the projection $( \, )_{-}$.
\vskip.2cm

If the $L_n$ are of the form
\be
    L_n = (L^n)_{+} \, ,    \label{L_n-GD}
\ee
we call (\ref{Lax-hier}) a \emph{Gelfand-Dickey-type hierarchy}.
In this case, a well-known argument (see \cite{Dick03}, for example) shows
that the integrability conditions (\ref{zero_curv}) of (\ref{Lax-hier})
are satisfied as a consequence of (\ref{Lax-hier}), so that no further
equations have to be added to those already given by (\ref{Lax-hier}).
\vskip.2cm

In the following we derive a very simple formula for the $E_n$.
Let us introduce $\tE_0 = 1$, $\tE_1 = -L$, and
\be
    \tE_{n+1} = (\tE_n)_{-} \, L  \qquad\quad  n=1,2, \ldots \; .
    \label{tE-recurs}
\ee

\begin{lemma}
\be
    \tE_n = - \sum_{k=1}^n (\tE_{n-k})_{+} \, L^k
    \qquad\quad  n=1,2, \ldots \; .
             \label{tE_recurs}
\ee
\end{lemma}
{\it Proof:}
Using
\bez
    (\tE_{n-k})_{+} \, L^k
  = \tE_{n-k} \, L^k - (\tE_{n-k})_{-} \, L^k
  = \tE_{n-k} \, L^k - \tE_{n-k+1} \, L^{k-1}
\eez
for $k=1, \ldots, n-1$, we have the following telescoping sum,
\bez
     \sum_{k=1}^{n-1} (\tE_{n-k})_{+} \, L^k
  = \sum_{k=1}^{n-1} \tE_{n-k} \, L^k
     - \sum_{k=0}^{n-2} \tE_{n-k} \, L^k
  = - \tE_n + \tE_1 \, L^{n-1}
  =  - \tE_n - L^n  \; .
\eez
\hfill $\blacksquare$

It is convenient to introduce the product (see also
\cite{EGK04Spitzer,DMH05KPalgebra}, for example)
\be
     X \vartriangle Y := (X)_{+} \, Y - X \, (Y)_{-}
                       = (X)_{+} \, (Y)_{+} - (X)_{-} \, (Y)_{-}
		       \label{vartriangle}
\ee
for $X,Y \in \Rf$.

\begin{lemma}
As a consequence of the hierarchy (\ref{Lax-hier}), we have
\be
    n \, \tE_n = - L^n - \sum_{k=1}^{n-1}
    \Big( \pa_{t_k}(\tE_{n-k}) + \tE_{n-k} \vartriangle L^k \Big)
    \qquad n=1,2, \ldots \; .
    \label{ntEn}
\ee
\end{lemma}
{\it Proof:} by induction on $n$. The formula trivially holds for $n=1$ and
is easily verified for $n=2$ using $L_{t_1} = [ (L)_{+} , L]$.
Let us assume that it holds for $n$.
Then
\bez
      n \, \tE_{n+1}
    = n \, (\tE_n)_{-} \, L
    = - (L^n)_{-} \, L - \sum_{k=1}^{n-1}
    \Big( \pa_{t_k}(\tE_{n-k})_{-} \, L - (\tE_{n-k})_{-} \, (L^k)_{-} \, L \Big)
\eez
by use of the induction hypothesis. With the help of
\bez
      \pa_{t_k}(\tE_{n-k})_{-} \, L
 &=& \pa_{t_k}(\tE_{n+1-k}) - (\tE_{n-k})_{-} \, \pa_{t_k}(L)
  = \pa_{t_k}(\tE_{n+1-k}) + (\tE_{n-k})_{-} \, [ (L^k)_{-} , L ] \\
 &=& \pa_{t_k}(\tE_{n+1-k}) + (\tE_{n-k})_{-} \, (L^k)_{-} \, L
     - \tE_{n+1-k} \, (L^k)_{-} \, ,
\eez
we obtain
\bez
      n \, \tE_{n+1}
  &=& - (L^n)_{-} \, L - \sum_{k=1}^{n-1}
        \Big( \pa_{t_k}(\tE_{n+1-k}) - \tE_{n+1-k} \, (L^k)_{-} \Big) \\
  &=& - \sum_{k=1}^n
      \Big( \pa_{t_k}(\tE_{n+1-k}) - \tE_{n+1-k} \, (L^k)_{-} \Big)
\eez
where we made again use of (\ref{Lax-hier}).
Finally we take account of
\bez
 & &   \sum_{k=1}^n \tE_{n+1-k} \, (L^k)_{-}
  = - \sum_{k=1}^n \tE_{n+1-k} \vartriangle L^k
     + \sum_{k=1}^n (\tE_{n+1-k})_{+} \, L^k   \\
 &=& - \sum_{k=1}^n \tE_{n+1-k} \vartriangle L^k
     + \sum_{k=1}^{n+1} (\tE_{n+1-k})_{+} \, L^k
     - (\tE_0)_{+} \, L^{n+1}    \\
 &=& - \sum_{k=1}^n \tE_{n+1-k} \vartriangle L^k
     - \tE_{n+1} - L^{n+1} \, ,
\eez
where we applied (\ref{tE_recurs}) in the last step,
to obtain (\ref{ntEn}) for $n+1$.
\hfill $\blacksquare$

\begin{theorem}
\label{theorem:E_n}
\be
    E_n = (\tE_n)_{+}   \qquad n=0,1,2, \ldots \; .
\ee
\end{theorem}
{\it Proof:} This is clearly true for $n=0,1$.
Taking the $(\ )_{+}$ part of (\ref{ntEn}), leads to
\bez
    n \, (\tE_n)_{+}
  = - \sum_{k=1}^n \Big( \pa_{t_k}(\tE_{n-k})_{+}
       + (\tE_{n-k})_{+} \, (L^k)_{+} \Big) \; .
\eez
Now our assertion follows by comparison with the recursion relation
(\ref{HnEn-recurs}) for the $E_n$.
\hfill $\blacksquare$

\section{A class of examples}
\label{section:examples}
\setcounter{equation}{0}
In this section we specialize the very general setting of the
previous section in order to make contact with some known hierarchies.
Our basic assumptions are formulated in the first subsection below.
An important tool is the notion of residue exploited in
section~\ref{subsec:res}. With its help we derive a general
`functional representation' in section~\ref{subsec:funct},
which also presents several examples.

\subsection{Preliminaries}
\label{subsec:ex_prel}
Let $\A$ be a unital associative algebra and $D$ an invertible linear operator
on $\A$ such that \\
(1) all its powers $D^n$, $n \in \mathbb{Z}$, are linearly independent (in the sense of a left $\A$-module), \\
(2) for all $a \in \A$,
\be
    D \, a = \Theta(a) \, D + \vartheta(a) \; .
\ee
Then $\Theta : \A \rightarrow \A$ has to be an algebra endomorphism and
$\vartheta$ a $\Theta$-twisted derivation,
\be
    \vartheta(a \, b) = \vartheta(a) \, b + \Theta(a) \, \vartheta(b) \; .
\ee
(3) $D$ and $\Theta$ are invertible (hence $\Theta$ is an automorphism of $\A$). \\
(4) $D$ commutes with all partial derivatives with respect to a set of coordinates,
say $t_n$, $n \in \mathbb{N}$.
This implies that also $\Theta$ and $\vartheta$ commute with all these
partial derivatives.
\vskip.2cm

As a consequence of conditions (2) and (3), we have
\be
    a \, D^{-1} = D^{-1} \, \Theta(a) + D^{-1} \, \vartheta(a) \, D^{-1}
\ee
and thus
\be
      D^{-1} \, a
  &=& \Theta^{-1}(a) \, D^{-1}
      - D^{-1} \, \vartheta(\Theta^{-1}(a)) \, D^{-1} \nonumber \\
  &=& \Theta^{-1}(a) \, D^{-1}
      - \Theta^{-1}( \vartheta(\Theta^{-1}(a))) \, D^{-2}
      + D^{-1} (\vartheta \circ \Theta^{-1})^2(a) \, D^{-2} \; .
\ee
Iteration leads to
\be
     D^{-1} a
 &=& \Theta^{-1}(a) \, D^{-1} - \vartheta_1(a) \, D^{-2}
     + \vartheta_2(a) \, D^{-3} - \ldots \nonumber \\
 &=& \Theta^{-1}(a) \, D^{-1} + \sum_{1 \leq n} (-1)^n \, \vartheta_n(a)
     \, D^{-n-1}   \qquad \quad \forall a \in \A
\ee
where
\be
   \vartheta_n := \Theta^{-1} \circ (\vartheta \circ \Theta^{-1})^n
   \qquad \quad    n=1,2,\ldots \; .
\ee
\vskip.2cm

\noindent
{\it Examples.} Let $\A$ be an algebra of matrices of functions. \\
1. Let $D$ be the operator of multiplication by a parameter $\zeta$.
The integer powers of $\zeta$ are linearly independent and commute with all $a \in \A$.
We have $\Theta = \id$ and $\vartheta = 0$. \\
2. $D = \pa$, the operator of partial differentiation with respect to a variable $x$. Then $\pa \, a = a_x + a \, \pa$ and
$\pa^{-1} a = a \, \pa^{-1} - a_x \, \pa^{-2} + a_{xx} \, \pa^{-3} - \ldots$,
where an index $x$ indicates a partial derivative with respect to $x$.
Here we have $\Theta = \id$ and $\vartheta = \pa_x$. \\
3. $D=\Lambda$, the shift operator $(\Lambda a)(x)=a(x+1)$ acting on a function $a$ (or a matrix of functions) of a variable $x$. Then
$\Lambda \, a = a^+ \, \Lambda$ and $\Lambda^{-1} a = a^{-} \, \Lambda^{-1}$ where $a^{\pm}(x) = a(x \pm 1)$.
In this case, $\Theta = \Lambda$ and $\vartheta = 0$. \\
4. $D = \Lambda_q$, where $q \not\in \{0,1 \}$ and $(\Lambda_q a)(x)=a(q \, x)$ acting on a function $a$ (or a matrix of functions)
of a variable $x$. Here we have $\Theta = \Lambda_q$ and $\vartheta = 0$.    \\
5. Let $D$ be the $q$-derivative operator
\be
    (\pa_q a)(x) = \frac{a(q \, x) - a(x)}{x \, (q-1)}  \label{q-deriv}
\ee
acting on functions of a variable $x$. In this case $\vartheta$ is
the $q$-derivative, and $\Theta = \Lambda_q$ with the $q$-shift operator defined above.
\hfill $\blacksquare$
\bigskip

Let $u_k \in \A$ and $\Rf$ be the algebra generated by the formal series
\be
    L = W \, u_0 \, D \, W^{-1}
      = u_0 \, D + u_1 + u_2 \, D^{-1} + u_3 \, D^{-2} + \ldots \; .
             \label{L-D}
\ee
We assume $u_0 \neq 0$ and choose the projections
\be
          (X)_{-} = X_{<0} \, , \qquad
	  (X)_{+} = X_{\geq 0}
\ee
of an element $X \in \Rf$ to its parts containing only negative, respectively
non-negative, powers of $D$. Another choice would be
$(X)_{-} = X_{<1}$, $(X)_{+} = X_{\geq 1}$.
This can be treated analogously and leads to further examples, see also section~\ref{section:mKP}.
As a consequence of our assumptions for the operator $D$, we have
$\Rf_{+} \, \Rf_{+} \subset \Rf_{+}$ and
$\Rf_{-} \, \Rf_{-} \subset \Rf_{-}$, as required in section~\ref{subsec:GD}.
In the following we consider Gelfand-Dickey-type hierarchies in this
specialized framework.
\vskip.2cm

\noindent
{\it Remark.}
Generically, the set of zero curvature equations (\ref{zero_curv}),
with (\ref{L_n-GD}) and (\ref{L-D}), actually implies the Lax hierarchy
(\ref{Lax-hier}) and is then \emph{equivalent} to it.
The following argument is taken from \cite{Ueno+Taka84}.
Writing (\ref{zero_curv}) in the form
\be
    \pa_{t_n}(L^m) - [ (L^n)_{+} , L^m ]
  = \pa_{t_n}(L^m)_{-} + \pa_{t_m}(L^n)_{+} - [ (L^n)_+ , (L^m)_{-} ] \, ,
    \label{zero_curv_split}
\ee
we observe that, for fixed $n$, on the right hand side the order of powers of $D$
is bounded above by $n$, whereas on the left hand side it increases with $m$.
Suppose $X^{(n)} := \pa_{t_n}(L) - [ (L^n)_{+} , L ] \neq 0$. The left hand 
side of (\ref{zero_curv_split}) then takes the form
\bez
    \pa_{t_n}(L^m) - [ (L^n)_{+} , L^m ]
  = \sum_{k=0}^{m-1} L^k \, X^{(n)} \, L^{m-1-k}   \qquad \quad m \geq 1 \, ,
\eez
which, for sufficiently large $m$, contains terms with powers of $D$
greater than $n$, which leads to a contradiction (unless the coefficients
of all those terms vanish because of very special properties of $L$).
Hence $X^{(n)} = 0$ and thus $\pa_{t_n}(L) - [ (L^n)_{+} , L ] = 0$.
\hfill $\blacksquare$
\bigskip

As a consequence of (\ref{Lax-hier}) with (\ref{L_n-GD}), we have
\be
        \pa_{t_n}(u_0) = 0   \qquad \quad n=1,2,\ldots    \label{u_0-const}
\ee
so that $u_0$ has to be constant. Furthermore, the first hierarchy equation
in particular leads to
\be
    u_{1,x} = u_0 \, \Theta(u_2) - u_2 \, \Theta^{-1}(u_0)  \label{u_1x} \; .
                  \label{u_1x-Theta}
\ee
In the following, we will look at $u_2$ as our `primary object'.
Introducing a potential $\phi$ such that
\be
     u_2 = \phi_x  \, ,     \label{u_2=phi_x}
\ee
equation (\ref{u_1x}) becomes
\be
    u_1 = u_0 \, \Theta(\phi) - \phi \, \Theta^{-1}(u_0)   \label{u1-eq}
\ee
(up to addition of an arbitrary element of $\A$ independent of $x$, which we
set to zero).
\vskip.2cm

In the following we use the abbreviations
\be
     a^+ := \Theta(a) \, , \qquad
     a^- := \Theta^{-1}(a)
\ee
where $a \in \A$.

\subsection{Taking residues}
\label{subsec:res}
We define the residue $\res(X)$ of $X \in \Rf$ as the
\emph{left}-coefficient\footnote{This means that, before reading off the
coefficient, we have to commute all powers of $D$ to the right.
If $D$ is given by example 1 or 2 in subsection~\ref{subsec:ex_prel}, the
residue does not depend on the ordering, however.}
of $D^{-1}$. It follows that
\be
   (L \, X_{<0})_{\geq 0} = u_0 \, \Theta(\res(X))
   \, , \qquad
   (X_{<0} \, L)_{\geq 0} = \res(X) \, \Theta^{-1}(u_0)
   \; .   \label{res-Theta}
\ee
The zero curvature condition (\ref{zero_curv}) with (\ref{L_n-GD})
can be written as follows,
\be
   \pa_{t_n} (L^m)_{<0} - \pa_{t_m} (L^n)_{<0} = [ (L^m)_{<0},(L^n)_{<0}]
    \; .
\ee
Taking the residue leads to
\be
     \res(L^m)_{t_n} = \res(L^n)_{t_m}    \; .
\ee
Hence there is a $\phi \in \A$ such that
\be
    \phi_{t_n} = \res(L^n) \; .   \label{phi-res}
\ee
For $n=1$, this is (\ref{u_2=phi_x}).

\begin{lemma}
\label{lemma:res(H_n)}
As a consequence of the hierarchy (\ref{Lax-hier}), we have
\be
    \res(\tE_n) = \hhi_n(\phi)
       \qquad \quad  n=1,2, \ldots
     \label{res(tE_n)}
\ee
\end{lemma}
{\bf Proof:} First we note that
\bez
    \res( X \vartriangle Y )
  = \res( X_{\geq 0} Y_{\geq 0}) - \res( X_{<0} Y_{<0} )
\eez
vanishes for all $X,Y \in \Rf$, since the first residue on the right hand
side vanishes as a consequence of
$\Rf_{\geq 0} \, \Rf_{\geq 0} \subset \Rf_{\geq 0}$,
and the second vanishes because $X_{<0} Y_{<0}$ does not contain
higher than $-2$ powers of $D$ according to our assumptions for $D$.
Taking also (\ref{phi-res}) into account, the residue of (\ref{ntEn}) is
\bez
    n \, \res(\tE_n) = - \phi_{t_n}
     - \sum_{k=1}^{n-1} \pa_{t_k}(\res(\tE_{n-k})) \; .
\eez
Our assertion now follows by comparing this recursion formula with
(\ref{chi-Schur-ids}), since $\res(\tE_1) = - \res(L) = \hhi_1(\phi)$.
\hfill $\blacksquare$
\bigskip

\begin{lemma}
\be
  E(\la) = 1 - \la \, u_0 \, D - \la \, (u_0 \, \phi^+ - \phi_{-[\la]} \, u_0^-)
	      \; .    \label{E(la)-phi}
\ee
\end{lemma}
{\it Proof:}
As a consequence of theorem~\ref{theorem:E_n}, equation (\ref{tE-recurs}) and lemma~\ref{lemma:res(H_n)}, we have
\bez
   E_{n+1} &=& ((\tE_n)_{<0} \, L)_{\geq 0}
            = ((\tE_n)_{<0} \, u_0 \, D)_{\geq 0}
	    = ( \res(\tE_n) \, D^{-1} u_0 \, D)_{\geq 0}  \\
           &=& \res(\tE_n) \, u_0^-
	    = \hhi_n(\phi) \, u_0^-
\eez
for $n=1,2, \ldots$. Hence
\bez
  E(\la) = 1 - \la \, u_0 \, D - \la \, (\phi - \phi_{-[\la]}) \, u_0^-
              - \la \, u_1
\eez
from which our assertion follows by use of (\ref{u1-eq}).
\hfill $\blacksquare$
\vskip.2cm

\noindent
{\it Remark.} Note that (\ref{E(la)-phi}) is \emph{polynomial} in $D$.
If we express the linear system (\ref{linsys}) in the form
$\chi_n(W) = H_n \, W$ with $H_n \in \Rf$, instead of (\ref{linsys-E_n}),
the resulting relation $E(\la)_{[\la]} \, H(\la) = 1$
with $H(\la) = \sum_{n \geq 0} \la^n H_n$ implies that $H(\la)$ is
an \emph{infinite} formal power series in $D$. This is the reason why
we chose to work with $E(\la)$ instead of $H(\la)$.
\hfill $\blacksquare$

\subsection{Functional representations}
\label{subsec:funct}
The next result evaluates the equations (\ref{EE}) in the framework
under consideration. Since by construction they are equivalent to
the zero curvature equations (\ref{zero_curv}), according to the
remark in section~\ref{subsec:ex_prel} they are generically also
equivalent to the complete hierarchy.

\begin{theorem}
In the present context, (\ref{EE}) is equivalent to
\be
      u_0 \, \vartheta((\phi_{[\la_1]} - \phi_{[\la_2]}) \, u_0^-)
  &=& ({1 \over \la_1} - u_0 \, \phi^+_{[\la_1]} + \phi u_0^-)
     ({1 \over \la_2} - u_0 \, \phi^+_{[\la_1]+[\la_2]} + \phi_{[\la_1]} \, u_0^-)
         \nonumber \\
  & & -({1 \over \la_2} - u_0 \, \phi^+_{[\la_2]} + \phi \, u_0^-)
     ({1 \over \la_1} - u_0 \, \phi^+_{[\la_1]+[\la_2]} + \phi_{[\la_2]} \, u_0^-)
            \label{generalizedFay}
\ee
(which is antisymmetric in $\la_1,\la_2$).
\end{theorem}
{\it Proof:}
Let us write (\ref{E(la)-phi}) as
\bez
     - {1 \over \la} \, E(\la)
   = u_0 \, D - \omega(\la) \qquad \mbox{where} \quad
     \omega(\la) := {1 \over \la} - u_0 \, \phi^+ + \phi_{-[\la]} \, u_0^-
     \; .
\eez
Using this in (\ref{EE}), we obtain the following two equations,
\bez
      u_0 \, (\omega^+(\la_1) - \omega^+(\la_2))
  &=& ( \omega(\la_1)_{-[\la_2]} - \omega(\la_2)_{-[\la_1]} ) \, u_0 \, ,  \\
      u_0 \, \vartheta(\omega(\la_2) - \omega(\la_1))
  &=& \omega(\la_1)_{-[\la_2]} \, \omega(\la_2)
      - \omega(\la_2)_{-[\la_1]} \, \omega(\la_1) \; .
\eez
The first equation turns out to be an identity by use of the definition
of $\omega(\la)$. So we are left with the second equation which is
\bez
      u_0 \, \vartheta((\phi_{-[\la_2]} - \phi_{-[\la_1]}) \, u_0^-)
  &=& ({1 \over \la_1} - u_0 \, \phi^+_{-[\la_2]} + \phi_{-[\la_1]-[\la_2]} \, u_0^-)
    ({1 \over \la_2} - u_0 \, \phi^+ + \phi_{-[\la_2]} \, u_0^-)   \nonumber \\
  & & -({1 \over \la_2} - u_0 \, \phi^+_{-[\la_1]} + \phi_{-[\la_1]-[\la_2]} \, u_0^-)
    ({1 \over \la_1} - u_0 \, \phi^+ + \phi_{-[\la_1]} \, u_0^-) \; .
\eez
After a Miwa shift $\mathbf{t} \to \mathbf{t} + [\la_1] + [\la_2]$, this
becomes (\ref{generalizedFay}).
\hfill $\blacksquare$
\bigskip

To order $\la_2^0 \, \la_1^n$, (\ref{generalizedFay}) yields
\be
    \chi_{n+1}(u_1) - u_0 \, \chi_n(\phi^+_{t_1} - \vartheta(\phi \, u_0^-))
  = u_1 \, \chi_n(u_1) + \sum_{k=1}^{n-1} u_0 \, \chi_k(\phi^+) \, \chi_{n-k}(u_1)
    + [u_0 \, \chi_n(\phi^+) , u_1]  \, ,
    \label{nonkp}
\ee
and to order $\la_2^m \, \la_1^n$, $m,n \geq 1$,
\be
   \chi_{n+1}(\chi_m(\varphi)) - \chi_{m+1}(\chi_n(\varphi))
 = \sum_{k=1}^n \chi_k(\varphi) \, \chi_{n-k}(\chi_m(\varphi))
     - \sum_{k=1}^m \chi_k(\varphi) \, \chi_{m-k}(\chi_n(\varphi))
    \label{chiphi}
\ee
where we introduced
\be
    \varphi := \Theta^{-1}(u_0) \, \phi = u^-_0 \, \phi \; .    \label{varphi}
\ee
In particular, for $m=1,n=2$, we recover the potential KP equation
\be
    {1 \over 3} \varphi_{tx} - {1 \over 12} \varphi_{xxxx}
    - {1 \over 4} \varphi_{yy}
  = {1 \over 2} (\varphi_x \varphi_x)_x - {1 \over 2} [\varphi_x , \varphi_y]
\ee
where $x=t_1$, $y=t_2$, and $t=t_3$.
In fact, as expressed in the subsequent theorem, the equations (\ref{chiphi})
are actually equivalent to the whole (noncommutative) potential KP hierarchy
with the dependent variable (\ref{varphi}).

\begin{theorem}
\label{theorem:ncKPdiffFay}
The equations (\ref{generalizedFay}) imply
\be
     \Big( \la_1^{-1} - \la_2^{-1} + \varphi_{[\la_2]} - \varphi_{[\la_1]} \Big)_x
 &=& \Big( \la_1^{-1} - \la_2^{-1} + \varphi_{[\la_2]} - \varphi_{[\la_1]} \Big) \,
    (\varphi_{[\la_1]+[\la_2]} - \varphi_{[\la_1]} - \varphi_{[\la_2]} + \varphi)
      \nonumber \\
  & & - [\varphi_{[\la_1]}-\varphi , \varphi_{[\la_2]}-\varphi]  \, ,
                        \label{ncKP-diffFay}
\ee
which is equivalent to
\be
  \sum_{i,j,k=1}^3 \epsilon_{ijk} \, \Big( \la_i^{-1}
  ( \varphi_{[\la_i]} - \varphi) + \varphi \, \varphi_{[\la_i]} \Big)_{[\la_k]}
  = 0   \, ,                 \label{ncKP-Fay}
\ee
where $\la_1, \la_2, \la_3$ are independent parameters and $\epsilon_{ijk}$
is totally antisymmetric with $\epsilon_{123} = 1$.
\end{theorem}
{\it Proof:} By expansion of (\ref{ncKP-diffFay}) in $\la_1, \la_2$, one recovers
(\ref{chiphi}), which we derived from (\ref{generalizedFay}).
Summing (\ref{ncKP-diffFay}) three times with cyclically permuted parameters
$\la_1, \la_2, \la_3$, leads to
\bez
  &&  \la_1^{-1} \left( (\varphi_{[\la_1]} - \varphi)_{[\la_3]}
          - (\varphi_{[\la_1]} - \varphi)_{[\la_2]} \right)
     + \la_2^{-1} \left( (\varphi_{[\la_2]} - \varphi)_{[\la_1]}
          - (\varphi_{[\la_2]} - \varphi)_{[\la_3]} \right)    \\
  && + \la_3^{-1} \left( (\varphi_{[\la_3]} - \varphi)_{[\la_2]}
          - (\varphi_{[\la_3]} - \varphi)_{[\la_1]} \right)    \\
 &=& [ \varphi \, ( \varphi_{[\la_1]} - \varphi_{[\la_3]} ) ]_{[\la_2]}
     + [ \varphi \, ( \varphi_{[\la_2]} - \varphi_{[\la_1]} ) ]_{[\la_3]}
     + [ \varphi \, ( \varphi_{[\la_3]} - \varphi_{[\la_2]} ) ]_{[\la_1]}
\eez
which can be rearranged to (\ref{ncKP-Fay}).
The limit $\la_3 \to 0$ leads back to (\ref{ncKP-diffFay}).
\hfill $\blacksquare$
\bigskip

As shown in \cite{DMH06nahier}, (\ref{ncKP-diffFay}) is a `noncommutative' version
of the \emph{differential Fay identity} for the (potential) KP hierarchy (see \cite{Shio86,Adle+vanM92,Adle+vanM94,Adle+vanM99,Taka+Take95,Dick95,Dick+Stra96}, for example).\footnote{In the commutative case, setting $\varphi = \tau_x/\tau$
and integrating once, leads to the familiar differential Fay identity. The form
in which we wrote (\ref{ncKP-diffFay}) facilitates this calculation. }
Equation (\ref{ncKP-Fay}), which already appeared in \cite{Bogd+Kono98,Bogd99},
is then a `noncommutative' version of the \emph{algebraic Fay identity}.
Here we have shown that, expressed as above in terms of (\ref{varphi}),
these formulae apply universally to \emph{all} examples in the class considered
in this section!
\vskip.2cm

In the special case of the KP hierarchy, (\ref{ncKP-diffFay}) is actually
equivalent to the hierarchy equations. This is \emph{not} true in general.
Typically (\ref{generalizedFay}) contains equations beyond those given
by (\ref{ncKP-diffFay}), and these are given by (\ref{nonkp}).
\vskip.2cm

\noindent
{\it Remark.} The KP hierarchy equations in the form (\ref{chiphi})
are the integrability conditions of the linear system
\be
    \chi_{m+1}(f) = - f \, \chi_m(\varphi)   \qquad m=1,2,\ldots \; .
      \label{KP-linsys}
\ee
In fact, as a consequence of the latter we have (by use of (\ref{chi-HS}))
\be
    \chi_{n+1}(\chi_{m+1}(f))
  = f \, \Big( - \chi_{n+1}(\chi_m(\varphi))
    + \sum_{k=1}^n \chi_k(\varphi) \, \chi_{n-k}(\chi_m(\varphi)) \Big)
    - \chi_1(f) \, \chi_n(\chi_m(\varphi)) \, \; .
\ee
Antisymmetrization in $m,n$ yields (\ref{chiphi}).
\hfill $\blacksquare$

\subsubsection{KP hierarchy}
The usual KP hierarchy in the Gelfand-Dickey framework (see \cite{Dick03}, for example)
is obtained by choosing $u_0=1$, $u_1=0$, $D=\pa$, the operator of partial
differentiation with respect to $x=t_1$, so that
$\Theta=\mathrm{id}$ and $\vartheta=\pa_x$.
Then (\ref{generalizedFay}) becomes
\be
     - (\phi_{[\la_1]} - \phi_{[\la_2]} )_x
 &=& (\la_2^{-1} - \phi_{[\la_2]}+\phi) (\la_1^{-1} - \phi_{[\la_1]+[\la_2]}
     + \phi_{[\la_2]})  \nonumber \\
 & & - (\la_1^{-1} - \phi_{[\la_1]} + \phi) (\la_2^{-1} - \phi_{[\la_1]+[\la_2]}
     + \phi_{[\la_1]})
\ee
which, after some simple algebraic manipulations, yields (\ref{ncKP-diffFay}) (note
that $\varphi = \phi$ in this case). Hence, in this case (\ref{generalizedFay})
(and thus (\ref{EE})) reduces to (\ref{ncKP-diffFay}).

\subsubsection{Discrete KP and $q$-KP}
The choice of $D$ in example 3 in subsection~\ref{subsec:ex_prel} leads to
the discrete KP hierarchy \cite{Adle+vanM99,Dick99,Feli+Onga01}.
For the choices of $D$ in examples 4 and 5 in subsection~\ref{subsec:ex_prel},
the hierarchy has been called `Frenkel system' \cite{Fren96}
and `KLR system' \cite{KLR97}, respectively,
in \cite{Adle+vanM99}, where the authors proved that both are isomorphic
to the discrete KP hierarchy (see also \cite{Dick03}).\footnote{The authors
of \cite{KLR97} actually
amputated the $q$-derivative by dropping the argument $x$ in the denominator
of (\ref{q-deriv}). See also \cite{Hain+Ilie97,Tu98,Ilie99,Carr03,Taka05qmKP}
for work on $q$-deformed KP hierarchies.}
\vskip.2cm

In the following we concentrate on examples 3 and 4 of subsection~\ref{subsec:ex_prel},
which can be treated simultaneously. Then
$\vartheta=0$ and $a^+ = \Theta(a) = \Lambda a \Lambda^{-1}$ with
$(\Lambda a)(s)= a(s+1)$ or $(\Lambda a)(s) = a(qs)$.
Furthermore, we choose $u_0=1$.
Then (\ref{generalizedFay}) takes the form
\be
 (\la_2^{-1} - \phi^+_{[\la_2]} + \phi) (\la_1^{-1} - \phi^+_{[\la_1]+[\la_2]}
   + \phi_{[\la_2]})
 = (\la_1^{-1} - \phi^+_{[\la_1]}+\phi) (\la_2^{-1} - \phi^+_{[\la_1]+[\la_2]}
   + \phi_{[\la_1]}) \; .
\ee
In the limit $\la_2 \to 0$, this yields
\be
   (\la^{-1} - \phi^+_{[\la]} + \phi)_x
 = (\phi^+ - \phi)(\la^{-1} - \phi^+_{[\la]} + \phi)
   - (\la^{-1} - \phi^+_{[\la]} + \phi)(\phi^+_{[\la]} - \phi_{[\la]}) \; .
   \label{dKP-ncBT}
\ee
Since according to theorem~\ref{theorem:ncKPdiffFay}, $\phi$ and $\phi^+$
both have to satisfy the KP hierarchy equations, the last equation
should represent a \emph{B\"acklund transformation} of the KP hierarchy.
Let us momentarily turn to the case of a commutative algebra $\A$.
Setting $\phi = \tau_x/\tau$ with a function $\tau$, an integration leads to
\be
    \tau^+_{[\la],x} \, \tau  - \tau^+_{[\la]} \, \tau_x
  = \la^{-1} \tau^+_{[\la]} \, \tau + \beta \, \tau^+ \, \tau_{[\la]}
    \label{dKPhier-BT}
\ee
where $\beta$ is an arbitrary $x$-independent function. This equation
has a limit as $\la \to 0$ if
\be
    \beta = - \la^{-1} + \beta_0 + \beta_1 \, \la + \ldots
\ee
where $\beta_0, \beta_1, \ldots$ are arbitrary $x$-independent functions.
If the latter are all set to zero, (\ref{dKPhier-BT}) becomes
equation (0.20) in \cite{Adle+vanM99} after a Miwa shift
$\mathbf{t} \mapsto \mathbf{t} - [\la]$. Treating them as parameters,
however, we should recover from (\ref{dKPhier-BT}) auto-B\"acklund
transformations of the KP hierarchy.
\vskip.2cm

Returning to the `noncommutative' case, expansion of (\ref{dKP-ncBT})
in powers of $\la$ leads to
\be
     (\chi_{n+1} - \chi_n \chi_1)(\phi^+) - \chi_{n+1}(\phi)
 &=& (\phi^+ - \phi) \, \chi_n(\phi^+ - \phi)
     + \sum_{k=1}^{n-1} \chi_k(\phi^+) \, \chi_{n-k}(\phi^+ - \phi)
                   \nonumber \\
 & & + [ \chi_n(\phi^+) , \phi^+ - \phi ] \, .
\ee
 For $n=1$, this is
\be
    (\phi^+ - \phi)_y - (\phi^+ + \phi)_{xx}
 = 2 \, \phi^+_x \, (\phi^+ - \phi) - 2 \, (\phi^+ - \phi) \, \phi_x  \; .
\ee
In the `commutative' case with $\phi = \tau_x/\tau$, this becomes
\be
    (D_x^2 - D_y) \, \tau^+ \cdot \tau = \beta_0 \, \tau^+ \, \tau
\ee
in terms of Hirota derivatives $D_x, D_y$. This equation is a well-known
auto-B\"acklund transformation of the KP
equation \cite{Hiro+Sats78,Roge+Shad82,Hiro04}.

\subsubsection{AKNS}
\label{subsec:AKNS}
With the choices of example 1 in subsection~\ref{subsec:ex_prel},
(\ref{generalizedFay}) reads
\be
  & & (\la_2^{-1} - u_0 \, \phi_{[\la_2]} + \phi \, u_0)
    (\la_1^{-1} - u_0 \, \phi_{[\la_1]+[\la_2]} + \phi_{[\la_2]} \, u_0)
                \nonumber \\
  &=& (\la_1^{-1} - u_0 \, \phi_{[\la_1]} + \phi \, u_0)
      (\la_2^{-1} - u_0 \, \phi_{[\la_1]+[\la_2]} + \phi_{[\la_1]} \, u_0) \, .
               \label{genFay-B-AKNS}
\ee
Choosing moreover
\be
    u_0 = \left( \begin{array}{cc} 1 & 0 \\ 0 & 0 \end{array} \right)
          \, , \qquad
    \phi = \left( \begin{array}{cc} p & q \\ - r & p' \end{array} \right) \, ,
    \label{AKNS_u0_phi}
\ee
we obtain the following system,
\be
 0 &=&  ( \la_1^{-1} -  \la_2^{-1} + p_{[\la_2]} - p_{[\la_1]} ) ( p_{[\la_1]+[\la_2]}
       - p_{[\la_1]} - p_{[\la_2]} + p )   \nonumber \\
   & & + (q \, r)_{[\la_2]} - (q \, r)_{[\la_1]}
       - [ p_{[\la_1]} - p  ,  p_{[\la_2]} - p ]  \, ,  \label{AKNS-pla} \\
 0 &=& ( \la_1^{-1} -  \la_2^{-1} + p_{[\la_2]} - p_{[\la_1]} ) \, q_{[\la_1]+[\la_2]}
       + \la_2^{-1} \, q_{[\la_1]} - \la_1^{-1} \, q_{[\la_2]} \, ,
                    \label{AKNS-qla}  \\
 0 &=& \la_1^{-1} \, ( r_{[\la_1]} - r )
       - \la_2^{-1} \, ( r_{[\la_2]} - r )
       + r \, ( p_{[\la_1]} - p_{[\la_2]} )  \, ,   \label{AKNS-rla}
\ee
which leaves $p'$ undetermined. In the limit $\la_2 \to 0$, this system yields
\be
          (p_{[\la]} - p)_x
  &=& q \, r - (q \, r )_{[\la]}   \, ,  \label{AKNS-p-eq} \\
       q_{[\la]} - q
  &=& \la \, q_{[\la],x} + \la \, ( p_{[\la]} - p ) \, q_{[\la]}  \, ,
            \label{AKNS-q-eq}   \\
  r_{[\la]} - r &=& \la \, r_x - \la \, r \, ( p_{[\la]} - p )  \; .
            \label{AKNS-r-eq}
\ee
Multiplying (\ref{AKNS-q-eq}) by $r$ from the right, (\ref{AKNS-r-eq}) by
$q_{[\la]}$ from the left, adding the resulting equations and using (\ref{AKNS-p-eq}),
we find
\be
    \left( p_{[\la]} - p + \la \, q_{[\la]} \, r \right)_x
  = [p_{[\la]} - p + \la \, q_{[\la]} \, r , p_{[\la]} - p] \; .
\ee
Expanding this equation in powers of $\la$, a simple induction argument
shows that\footnote{Constants of integration are set to zero.}
\be
     p_{[\la]} - p = - \la \, q_{[\la]} \, r   \; .   \label{AKNS-p_la-p}
\ee
Eliminating $p$ from (\ref{AKNS-q-eq}) and (\ref{AKNS-r-eq}) with the help
of this formula, we arrive at
\be
    q_{-[\la]} - q + \la \, q_x = \la^2 \, q \, r_{-[\la]} \, q  \, ,  \qquad
     r_{[\la]} - r - \la \, r_x = \la^2 \, r \, q_{[\la]} \, r \, ,
     \label{AKNS-funct}
\ee
which is a `functional representation' of the AKNS hierarchy
\cite{Veks02,Prit+Veks02}, generalized to the case
where $q$ and $r$ are matrices with entries from any associative algebra.
Expanding the above system in powers of $\la$, leads in lowest order to
\be
   q_{t_2} = q_{xx} - 2 \, q \, r \, q  \, , \qquad
   r_{t_2} = - r_{xx} + 2 \, r \, q \, r  \; .   \label{AKNS-qr_t2}
\ee
To next order in $\la$ we obtain, after use of the first system,
\be
   q_{t_3} = q_{xxx} - 3 \, ( q \, r \, q_x + q_x \, r \, q ) \, , \qquad
   r_{t_3} = r_{xxx} - 3 \, ( r \, q \, r_x + r_x \, q \, r ) \; .
\ee
For example, we may choose $q$ and $r$ as $M \times N$ and $N \times M$
matrices, respectively, with entries from any, possibly noncommutative,
associative algebra.
In this way, (\ref{AKNS-qr_t2}) also covers the case of (coupled)
vector nonlinear Schr\"odinger equations
(if we replace $t_2$ by the imaginary variable $\imath \, t$).\footnote{See
also \cite{Ford+Kuli83,APT04} and references therein concerning
nonlinear Schr\"odinger-type equations.}
\bigskip

\noindent
{\it Remark.} The equations (\ref{AKNS-pla}), (\ref{AKNS-qla})
and (\ref{AKNS-rla}) should reduce completely to the system
(\ref{AKNS-funct}) of functional equations,
since it contains the full set of hierarchy equations.
We verify that this is indeed the case. First we note that
the equations (\ref{AKNS-funct}) imply
\be
    (q \, r)_{[\la]} - q \, r  = \la \, (q_{[\la]} \, r)_{x} \; .
\ee
Introducing $p$ such that $p_x = - q \, r$, this yields (\ref{AKNS-p_la-p})
after an $x$-integration.
Inserting (\ref{AKNS-p_la-p}) in (\ref{AKNS-rla}) turns it into
\be
      \la_1^{-1} ( r_{[\la_1]} - r ) + \la_1 \, r \, q_{[\la_1]} \, r
   =  \la_2^{-1} ( r_{[\la_2]} - r ) + \la_2 \, r \, q_{[\la_2]} \, r \, ,
\ee
which means that $\la^{-1} ( r_{[\la]} - r ) + \la \, r \, q_{[\la_1]} \, r$
is independent of $\la$. This is obviously equivalent to
the second of equations (\ref{AKNS-funct}).
Inserting (\ref{AKNS-p_la-p}) in (\ref{AKNS-qla}), transforms it into
\be
      \la_1^{-1} ( q_{[\la_1]} - q )_{[\la_2]}
    - \la_2^{-1} ( q_{[\la_2]} - q )_{[\la_1]}
    + ( \la_1 \, q_{[\la_1]} - \la_2 \, q_{[\la_2]} ) \, r \,  q_{[\la_1]+\la_2]}
      = 0   \, ,
\ee
which indeed holds as a consequence of the second of equations (\ref{AKNS-funct}),
and the integrability condition of (\ref{AKNS-p_la-p}), which is
\be
   \la_2^{-1} \left( ( q_{[\la_1]} \, r)_{[\la_2]} - q_{[\la_1]} \, r \right)
  - \la_1^{-1} \left( ( q_{[\la_2]} \, r)_{[\la_1]} - q_{[\la_2]} \, r \right)
  = 0   \; .
\ee
A lengthier calculation shows that also (\ref{AKNS-pla}) results from
(\ref{AKNS-funct}).
\hfill $\blacksquare$
\vskip.2cm

\noindent
{\it Remark.} Our general results imply that
\be
  \varphi = u_0 \, \phi
          = \left( \begin{array}{cc} p & q \\ 0 & 0 \end{array} \right) \, ,
\ee
where $p_x = - q \, r$ according to (\ref{AKNS-p_la-p}),
solves the KP hierarchy as a consequence of the AKNS hierarchy equations.
Inspection of (\ref{ncKP-diffFay}) then shows that $p$ satisfies the KP hierarchy.
\hfill $\blacksquare$
\vskip.2cm

\noindent
{\it Remark.} Since $u_0$ satisfies $u_0^2 = u_0$, the dressing relation
$L = W \, u_0 \, \zeta W^{-1}$ implies
$L^2 = W (u_0 \, \zeta)^2 W^{-1} = \zeta \, L$.
As a consequence, $L^n = \zeta^{n-1} \, L$, so that the Lax equations
(\ref{Lax-hier}) take a more familiar form of (generalized)
AKNS hierarchies \cite{Wils81,Drin+Soko84,DMH04hier}.
\hfill $\blacksquare$

\section{Modified KP hierarchy}
\label{section:mKP}
\setcounter{equation}{0}
In this section we derive a functional representation of the
\emph{modified KP hierarchy} \cite{Kupe89,Chen95,Gesz+Unte95,Kupe00},
which is given by
\be
    \pa_{t_n}(L) = [ (L^n)_{\geq 1} , L ]  \qquad n=1,2,\ldots
\ee
where
\be
    L = \pa + u_1 + u_2 \, \pa^{-1} + \cdots
\ee
with the operator $\pa$ of partial differentiation with respect to $x = t_1$
(cf example 2 in section~\ref{subsec:ex_prel}),
and coefficients from some associative algebra $\A$.
Obviously, the $E_n$, $n >0$, which are determined by theorem~\ref{theorem:E_n}
and the recursion relation (\ref{tE-recurs}), are linearly homogeneous in $\pa$.
Hence
\be
    E(\la) = 1 - \la \, \omega(\la) \, \pa \, ,
\ee
where
\be
    \omega(\la) = 1 + u_1 \, \la + (u_2 + u_1^2) \, \la^2 + \ldots
       \label{mKP-om(la)}
\ee
is a power series in $\la$ with coefficients in $\A$.
The equation (\ref{EE}) now takes the form
\be
    (1 - \la_1 \, \omega(\la_1)_{-[\la_2]} \, \pa)(1 - \la_2 \, \omega(\la_2) \, \pa)
  = (1 - \la_2 \, \omega(\la_2)_{-[\la_1]} \, \pa)(1 - \la_1 \, \omega(\la_1) \, \pa)
    \; .
\ee
Expansion in powers of $\pa$ leads to
\be
    \omega(\la_1)_{-[\la_2]} \, \omega(\la_2)
  = \omega(\la_2)_{-[\la_1]} \, \omega(\la_1)     \label{mkp1}
\ee
and
\be
    {1 \over \la_2}(\omega(\la_1) - \omega(\la_1)_{-[\la_2]})
  - {1 \over \la_1}(\omega(\la_2) - \omega(\la_2)_{-[\la_1]})
  = \omega(\la_2)_{-[\la_1]} \, \omega(\la_1)_x
    - \omega(\la_1)_{-[\la_2]} \, \omega(\la_2)_x \; .    \label{mkp2}
\ee
The first equation is solved by
\be
    \omega(\la) = f_{-[\la]} \, f^{-1}     \label{omf}
\ee
with some invertible element $f \in \A$.
By comparison with (\ref{mKP-om(la)}),
\be
    v := u_1 = - f_x \, f^{-1} \; .
\ee
Next we use (\ref{omf}) in (\ref{mkp2}), multiply by $(f^{-1})_{-[\la_1]-[\la_2]}$
from the left and by $f$ from the right, and apply a Miwa shift
$\mathbf{t} \to \mathbf{t} + [\la_1] + [\la_2]$ to obtain the following
functional representation of the mKP hierarchy,
\be
      \la_2^{-1} \left( (f^{-1} \, f_{[\la_2]})_{[\la_1]}
         - f^{-1} \, f_{[\la_2]} \right)
    - \la_1^{-1} \left( (f^{-1} \, f_{[\la_1]})_{[\la_2]}
         - f^{-1} \, f_{[\la_1]} \right)
   = (f^{-1} \, f_x)_{[\la_1]} - (f^{-1} \, f_x)_{[\la_2]} \, , \quad
        \label{mKP-diffFay}
\ee
which is a noncommutative version of the differential Fay identity
of the mKP hierarchy. Adding it three times with cyclically permuted
parameters $\la_1, \la_2, \la_3$, leads to the corresponding algebraic
Fay identity
\be
  &&  \la_1^{-1} \left( (f^{-1} \, f_{[\la_1]})_{[\la_3]}
        - (f^{-1} \, f_{[\la_1]})_{[\la_2]} \right)
     + \la_2^{-1} \left( (f^{-1} \, f_{[\la_2]})_{[\la_1]}
        - (f^{-1} \, f_{[\la_2]})_{[\la_3]} \right)  \nonumber \\
  && + \la_3^{-1} \left( (f^{-1} \, f_{[\la_3]})_{[\la_2]}
        - (f^{-1} \, f_{[\la_3]})_{[\la_1]} \right)
   = 0 \; .
\ee
(\ref{mKP-diffFay}) is equivalent to
\be
    \chi_n\left( f^{-1} \, \chi_{m+1}(f) \right)
  = \chi_m\left( f^{-1} \, \chi_{n+1}(f) \right)
    \qquad \quad  m,n =1,2, \ldots  \; .       \label{mKP-chi(f)}
\ee
For $m=1$ and $n=2$, this yields
\be
   (f_t - \frac{1}{4} \, f_{xxx})_x - \frac{3}{4} \, f_{yy}
   - f_x f^{-1} (f_t - f_{xxx})
   + \frac{3}{4} \Big( f_y f^{-1} + (f_x f^{-1})_x - (f_x f^{-1})^2 \Big)
     (f_{xx} + f_y) = 0 \quad
\ee
which, multiplied by $f^{-1}$ from the right, leads to
\be
   v_t - \frac{1}{4} \, v_{xxx} + \frac{3}{2} \, v \, v_x \, v
   - \frac{3}{4} \, [ v , v_{xx} ]
   - \frac{3}{4} \Big( w_y + w \, (v_x - v^2)
       + (v_x + v^2) \, w \Big) = 0  \; .   \label{ncmKP}
\ee
where we introduced
\be
        w := - f_y \, f^{-1}  \; .
\ee
Since $w_x = v_y - [v , w]$, there is no way to
express also the terms involving $w$ completely in terms of $v$, unless
we assume that $\A$ is commutative, in which case
we obtain the mKP equation (see \cite{Blas98}, for example)
\be
   v_t - \frac{1}{4} \, v_{xxx} + \frac{3}{2} \, v^2 v_x
    - \frac{3}{4} \, \pa^{-1}(v_{yy})
    - \frac{3}{2} \, v_x \, \pa^{-1}(v_y) = 0 \; .
\ee
Returning to the noncommutative case and setting $f_y =0$, (\ref{ncmKP})
reduces to the second version of a `matrix mKdV equation'
in \cite{Olve+Soko98assoc},
\be
   v_t - \frac{1}{4} \, v_{xxx} + \frac{3}{2} \, v \, v_x \, v
   - \frac{3}{4} \, [ v , v_{xx} ] = 0  \; .
\ee
\vskip.2cm

The mKP hierarchy in the form (\ref{mKP-chi(f)})
implies the existence of a potential $\phi$ such that
\be
    f^{-1} \chi_{n+1}(f) = - \chi_n(\phi) \qquad \quad n=1,2,\ldots \, ,
\ee
which can, of course, also be written as
\be
    \chi_{n+1}(f) = - f \, \chi_n(\phi)   \qquad \quad n=1,2,\ldots \; .
    \label{mKP-KPlin}
\ee
Whereas the first form of this linear system naturally imposes
integrability conditions on $f$, namely the mKP equations in the form
(\ref{mKP-chi(f)}), the second gives rise to integrability conditions
in terms of the potential $\phi$. Since (\ref{mKP-KPlin}) coincides with
(\ref{KP-linsys}), the latter are precisely the potential KP hierarchy
equations in the form (\ref{chiphi}) (with $\varphi$ replaced by $\phi$).
\vskip.2cm

Furthermore, the above linear system mediates between the two hierarchies.
The lowest ($n=1$) member of (\ref{mKP-KPlin}) reads
\be
  u := \phi_x = - f^{-1} \chi_2(f) = - {1 \over 2}f^{-1} (f_y + f_{xx}) \; .
\ee
In the `commutative' case, we can express the right hand side in terms
of $v$ and recover the Miura transformation (see \cite{Dubr+Kono91,Bogd99},
for example), which maps solutions of the mKP to solutions of the KP equation.
\vskip.2cm

\noindent
{\it Remark.} The `duality' between the mKP and the KP hierarchy, which
emerged here, reminds us of the relation between different forms of the
(anti-) self-dual Yang-Mills equation (see \cite{Maso+Wood96}, for example),
and also the analogous relation between the principal chiral model
and its pseudo-dual. In the latter case, the analog of (\ref{mKP-KPlin}) is
\be
   \pa_{t_{n+1}}(f) = - f \, \pa_{t_n}(\phi)   \qquad \quad n=1,2,\ldots
       \label{c-pc-lin}
\ee
which gives rise to the following two versions of integrability conditions,
\be
    \pa_{t_n}( f^{-1} \pa_{t_{m+1}}(f)) = \pa_{t_m}( f^{-1} \pa_{t_{n+1}}(f))
\ee
(the analog of the mKP hierarchy equations in the form (\ref{mKP-chi(f)})) and
\be
    \pa_{t_{n+1}} \pa_{t_m}(\phi) - \pa_{t_{m+1}} \pa_{t_n}(\phi)
  = [\pa_{t_n}(\phi) , \pa_{t_m}(\phi)]
\ee
(the analog of the KP hierarchy equations in the form obtained from
(\ref{mKP-KPlin})). The condition $t_{n+2}=t_n$ reduces these systems to the
principal chiral model equation
\be
    \pa_{t_2}( f^{-1} \pa_{t_2}(f)) = \pa_{t_1}( f^{-1} \pa_{t_1}(f)) \, ,
\ee
respectively the pseudodual chiral model equation \cite{Curt+Zach94,Curt+Zach95para}
\be
    \pa_{t_1}\pa_{t_1}(\phi) - \pa_{t_2}\pa_{t_2}(\phi)
  = [\pa_{t_2}(\phi) , \pa_{t_1}(\phi)]  \; .
\ee
Both are known to be expressions of the system
\be
    F := \d A + A \wedge A = 0 \, , \qquad
    \d \star A =0 \, ,
\ee
for a 1-form $A$ in two dimensions, where $\star$ is the Euclidean Hodge operator.
The chiral model results by solving the first equation by $A = f^{-1} \, \d f$
with an invertible matrix $f$. The pseudodual model is obtained by solving
the second equation via $A = \star \, \d \phi$ with a matrix of functions $\phi$.
As we have seen, both models also emerge from the linear system (\ref{c-pc-lin}).
The latter has the following generalization,
\be
    \pa_{s_n}(f) = - f \, \pa_{t_n}(\phi)    \label{sdYM-lin}
\ee
with an additional set of independent variables $s_n$, $n=1,2,\ldots$.
The corresponding integrability conditions are
\be
    \pa_{t_n}( f^{-1} \pa_{s_m}(f)) = \pa_{t_m}( f^{-1} \pa_{s_n}(f)) \, ,
        \label{sdYMhier1}
\ee
respectively
\be
   \pa_{s_n} \pa_{t_m}(\phi) - \pa_{s_m} \pa_{t_n}(\phi)
     = [\pa_{t_n}(\phi) , \pa_{t_m}(\phi)] \; .    \label{sdYMhier2}
\ee
(\ref{sdYMhier1}) represents a self-dual Yang-Mills hierarchy \cite{Beal+Coif86,Naka88,Naka91,ACT93,Maso+Wood96}, and (\ref{sdYMhier2})
is its `dual' version.
It remains to be seen whether there is a meaningful `reverse'
analog of (\ref{sdYM-lin}) which generalizes (\ref{mKP-KPlin}) in a similar
way as (\ref{sdYM-lin}) generalizes (\ref{c-pc-lin}).
\hfill $\blacksquare$

\section{Conclusions}
\label{section:concl}
\setcounter{equation}{0}
In this work we formulated a rather general approach towards
`functional representations' of integrable hierarchies, in particular
analogs of `Fay identities'. This formalism is
\emph{not} restricted to commutative dependent variables, but
genuinely applies to `noncommutative hierarchies', where the dependent
variables live in any noncommutative algebra\footnote{This includes
the case of an algebra of functions supplied with a Moyal-product,
for example.}, like matrix KP hierarchies.
The central part of the formalism is general enough to embrace
many more integrable hierarchies and should serve to unify individual
results in the literature. We provided corresponding examples, but
by far did not exhaust the possibilities.
\vskip.2cm

Apart from the fact that our approach presents a fairly simple and
systematic way towards functional representations of specific
`noncommutative' integrable hierarchies, it also serves beyond that
as a tool in `integrable hierarchy theory'. For example,
the application of the formalism in section~\ref{section:mKP}
nicely displays the `duality' between the mKP and the KP hierarchy.

\addcontentsline{toc}{section}{\numberline{}References}

\end{document}